\documentclass{aastex}
\usepackage{spr-astr-addons}
\usepackage{url}
\usepackage{color}

\RequirePackage{color}

\begin{document}

\title{Plasma Magnetosphere Formation Around Oscillating Magnetized Neutron Stars}
\slugcomment{}
%% Running heads
\shorttitle{}
\shortauthors{Ahmedov et al.}

\author{B.J. Ahmedov\altaffilmark{1,2,3}} \and \author{V.S. Morozova\altaffilmark{1,3}}

%\email{\emaila}

\altaffiltext{1}{Ulugh Begh Astronomical Institute,
Astronomicheskaya 33, Tashkent 100052, Uzbekistan}
\altaffiltext{2}{Institute of Nuclear Physics,
        Ulughbek, Tashkent 100214, Uzbekistan}
\altaffiltext{3}{International Centre for Theoretical Physics,
Strada Costiera 11, 34014 Trieste, Italy}

\begin{abstract}
The notion of death line of rotating pulsars is applied to model
of oscillating neutron stars. It is shown that the magnetosphere
of typical non-rotating oscillating stars may not contain
secondary plasma to support the generation of radio emission in
the region of open field lines of plasma magnetosphere.
\end{abstract}

\keywords{Death line; plasma magnetosphere; oscillating neutron
stars.}

\section{Introduction}

The model describing formation of the magnetosphere around highly
magnetized rotating neutron star was first proposed by
\cite{gj69}. They argued that rotating magnetized neutron star
cannot be surrounded by vacuum. Electric field, generated on the
surface of the star due to rotation, pulls particles out of the
surface and accelerates them along open magnetic field lines.
Accelerated particles emit gamma rays due to curvature of magnetic
field lines. Gamma rays then produce electron-positron pairs,
which in turn accelerate and produce photons again. There arises a
chain reaction of multiplication of electrons, positrons and
gamma-ray quanta near the surface of polar cap which leads to the
formation of the plasma magnetosphere.

There are, however, certain conditions limiting the values of the
pulsars rotation period $P$ and the strength of magnetic field $B$
under which formation of the magnetosphere can be realized. Death
line is the $P-\dot{P}$ (or $P-B$, or $P-\Phi$) diagram which
indicates the region where pulsar can support radio emission from
magnetosphere (here $\Phi$ is the potential of accelerating field,
$\dot{P}$ means the time derivative of a period). The essence of
death line is that beyond certain conditions on the period and
surface magnetic field strength of the pulsar sufficient amount of
secondary plasma which is responsible for radio emission cannot be
generated. In this case either potential drop cannot accelerate
the primary particles along magnetic field lines enough to produce
high energy gamma rays, or the component of magnetic field
perpendicular to gamma ray propagation is not high enough to
produce the pair. The mechanism of pair creation and pulsar radio
emission is thoroughly described in the papers of \cite{Er66},
\cite{Manch80}, \cite{RS75}, \cite{Med07}.

Since its appearance in 1975 in the work of \cite{RS75} the notion
of death line has been reviewed and revised by many authors (see
e.g. \cite{AS79}, \cite{CR93}, \cite{RR94}). Numerical approach to
the calculation of death line for the case of dipolar magnetic
field of rotating neutron star is presented by \cite{MusHar02}. Up
to now the notion of death line is satisfied well enough and all
known radio pulsars are located in the region, which is predicted
by theory to be "radio-loud", i. e. capable to produce radio
emission (see \cite{Qiao03}, \cite{ZhHar00}).

Investigation of death line may help in better understanding the
mechanism of radio emission from neutron stars; it can serve as a
tool for checking various models of pulsar radiation. The works of
\cite{ZhHar00}, \cite{ZhHar99} are devoted to exploring the model
dependence of the pulsar death lines where two different types of
acceleration models are investigated in detail.

In our research we will study the death line's formalism for
oscillating non-rotating neutron stars by using approach developed
by \cite{Kan04}. Our intention is to show that in the framework of
present concepts of death line of pulsar one can predict that
oscillating non-rotating typical neutron stars do not produce
enough plasma in the region of open field lines to generate radio
emission and, therefore, they are radio-quiet. However the
oscillating star may have plasma magnetosphere formed in earlier
stages of stellar evolution when it was rotating. According
to~\cite{Jones86} the cohesive energy of the particles on surface
of a pulsar is small to prevent them escape into the magnetosphere
under the vacuum electric field. But theory of pulsar activity
tells that for producing radio emission there should be continuous
processes of plasma formation maintained in the vicinity of the
neutron star's polar cap. This condition is not satisfied on
investigated stage of neutron star's evolution.

\section{Death line application to oscillating stars}

As it was shown in a number of papers (see \cite{RS75},
\cite{Qiao03}, \cite{Manch80}) the production of secondary plasma
in the open field lines region of the pulsar may be realized if
the potential accelerating the primary particle is large enough
(for the inverse Compton scattering of thermal photons the Lorentz
factor of primary particles must rich $\gamma>mc^2/2kT$, where
$mc^2$ is the particle's rest mass, $T$ is the temperature of the
beam, $c$ is the speed of light) and if the perpendicular to
$\gamma$-ray component of magnetic field gets the value
\begin{equation}
B_{\bot}=B_c\frac{0.2mc^2}{\hbar\omega}\ .
\end{equation}

Here $\hbar\omega$ is the energy of photon emitted by the primary
particle and $B_c\equiv m^2 c^3/e\hbar=4.414\times 10^{13}$G is
the natural quantum mechanical unit for magnetic field strength
($e$ is the charge of electron). The conditions for pair
production are better satisfied at the point $\eta_0=1.5\eta$,
where $\eta$ is the distance from the stellar center (in units of
stellar radius) to the place where the gamma ray is emitted and
$\eta_0$ is the place where the pair is arising. Below this
condition the angle between magnetic field and $\gamma$-ray is
small, up to this condition the strength of magnetic field is low.
Using this assumption \cite{Kan04} got the expression for the
death line of radiopulsars in the following form
\begin{equation}
\frac{10^{15}(1.5\eta)^3P^{1/2}}{B_{12}\gamma mc^2}=\frac{\eta}{2}R\
,
\end{equation}
where $B_{12}$ is the strength of pulsar magnetic field, normed on
$10^{12} {\rm G}$, $R$ is the radius of the star.

The accelerating electric field in the region of the open field
lines of the pulsar $E\sim\Omega R B/c$, while corresponding
potential drop $\Phi\sim\Omega B R^2/c$.

Consider now non-rotating neutron star oscillating with the
frequency $\omega_{\rm osc}$. Oscillations of the star are assumed
to be toroidal, what means that the components of the oscillation
velocity in spherical coordinates $(r,\theta,\phi)$ have the
following form
\begin{eqnarray}
\delta v^{\hat{r}}&=&0 \ , \\
\delta v^{\hat{\theta}}&=&e^{-i\omega_{\rm o s
c}t}\eta(r)\frac{1}{\sin\theta}\partial_{\phi}Y_{l'm'}(\theta,\phi)
\ , \nonumber \\
\delta v^{\hat{\phi}}&=&-e^{-i\omega_{\rm o s
c}t}\eta(r)\partial_{\theta}Y_{l'm'}(\theta,\phi) \ . \nonumber
\end{eqnarray}
Here $\eta(r)$ is the transversal velocity amplitude, oscillation
are assumed to be small-amplitude and the magnetic field of the
star has dipolar configuration.

Oscillations of the star produce electric field which in analogy
with the pulsar's accelerating electric field can be written as
\begin{equation}
\label{osc} E_{osc}\sim \frac{\omega_{\rm osc}\xi}{c}B =
\frac{\omega_{\rm osc}R}{c}\left(\frac{\xi}{R}\right) B =
\frac{\omega_{\rm eff}R}{c}B\ .
\end{equation}

The effective frequency of stellar pulsations will be $\omega_{\rm
eff}\sim\xi\omega_{\rm osc}/R$, where $\xi$ is the amplitude of
oscillations.

The physical processes in the magnetosphere of oscillating neutron
stars were thoroughly studied by Timokhin et. al. (2000). In
particular, it was discussed an important question on formation of
the region of open magnetic field lines in the magnetosphere of an
oscillating star. It was shown that for every mode of oscillations
the angle (on the surface of the star) of the last open field line
$\theta_0$ should be determined self-consistently using equation
for the Alfvenic surface. It was found that for oscillation modes
with $m<3$ the angle $\theta_0$ is small and expression for the
Goldreich-Julian charge density in the approximation of small
angles $\theta$ may be written as
\begin{equation}
\rho_{GJ}\sim\frac{B\eta}{R c}\cdot\theta^m\ .
\end{equation}

Accelerating scalar potential above the surface of the pulsar,
which can be found by means of Poisson equation
\begin{equation}
\nabla \cdot (\frac{1}{N}\nabla\Phi)= -4\pi(\rho-\rho_{GJ}) \ ,
\end{equation}
where $N\equiv(1-2M/r)^{1/2}$ is the gravitational lapse function
and $\rho-\rho_{GJ}$ is the effective space charge density, is
proportional to $\theta_0^2$ for rotating neutron star. In the
case of oscillating magnetized neutron star one can find by
introducing dimensionless angular variable $\theta/\theta_0$ and
taking into account angular dependence of $\rho_{GJ}$ that scalar
potential $\Phi\sim\theta_0^{m+2}$. Consequently, for oscillation
modes with $m<3$ accelerating scalar potential will contain
additional small parameter.

Taking into account that oscillation amplitude $\xi$ is about
$10^3$ times smaller than radius of star $R$ and assuming that
$\theta_0\sim 10^{-2}$ one can find that potential drop
accelerating the primary particles for modes with $m<3$ will be
$10^3\times10^{(2m)}$ times smaller with compare to that in the
case of rotating neutron stars. The result is that this potential
will not be large enough to provide the primary particle with
sufficient Lorentz factor to emit gamma ray which would be
possible to generate a pair.

\begin{figure}
\includegraphics[width=0.48\textwidth]{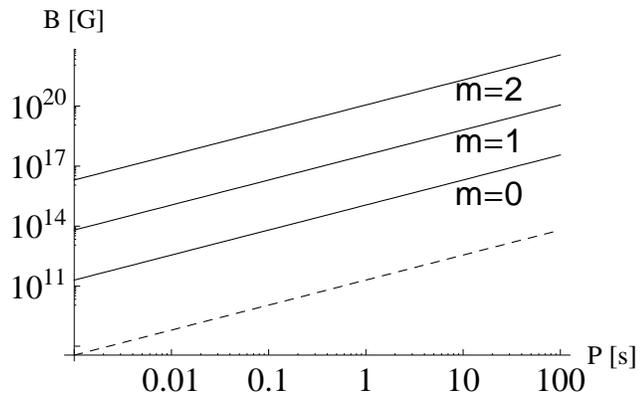}
\caption{Death lines for rotating radiopulsar (dashed) and for
oscillating non-rotating neutron star (solid).} \label{dl}
\end{figure}

Figure \ref{dl} illustrates the situation. Dashed curve is the
death line for the rotating radiopulsar in frames of the
space-charge-limited flow model, adopted from the paper
\cite{Kan04}. Solid line presents the death line for non-rotating
oscillating neutron star. It is plotted for real (not effective)
values of pulsation periods, which are $\sim 1000$ times shorter
than $P_{\rm eff}$. Using the graph one can see that radioemission
from non-rotating stars may be observed only if the frequency of
oscillations is rather large. Typical oscillating non-rotating
stars with parameters $\omega_{\rm osc}<0.5~ {\rm kHz}$ and
surface magnetic field $B\sim 10^{12}~ {\rm G}$ will lie beyond
the radio-loud region. It means that plasma may not be produced in
the region of open field lines of oscillating non-rotating neutron
star's magnetosphere with typical parameters. It means that vacuum
electrodynamic model of oscillating stars (\cite{mshzh84},
\cite{mt86}, \cite{ra04}) is more preferable rather than plasma
magnetosphere model (\cite{tbs00}).

\section{Conclusions}

Observations of radio emission from a star's environment are
possible only in the case if the magnetosphere of the star is
filled with enough amount of secondary plasma. Primary particles,
extracted from the surface of the star, generate secondary plasma
oscillations resulting in radio photons emission. As it is seen
from our results, typical non-rotating oscillating neutron star
has not enough conditions to produce secondary plasma and probably
may not produce plasma magnetosphere. However magnetars
(\cite{d98}) with the high surface magnetic fields of order
$10^{14}{\rm G}$ are above solid line from figure \ref{dl}.
Furthermore, there is now observational evidence for stellar
oscillations coming from the observation of quasi-periodic
oscillations (QPOs) following giant flares of soft gamma-ray
repeaters (SGRs) (\cite{ietal05}, \cite{sw05}, \cite{ws06},
\cite{ws07a}, \cite{ws07b}). The analysis of the X-ray data from
SGRs has in fact revealed that the decaying part of spectrum
exhibits a number of oscillations with frequencies in the range of
a few tenths of ${\rm Hz}$ to a few hundred ${\rm Hz}$ that agree
reasonably well with the expected toroidal modes of the magnetar
crust (\cite{d98}). Due to this one may expect formation of plasma
magnetosphere around oscillating relativistic stars with stronger
magnetic fields as magnetars.

\section*{Acknowledgments}

This research is supported in part by the UzFFR (projects 5-08 and
29-08) and projects FA-F1-079, FA-F2-F061 and A13-226 of the UzAS,
by the ICTP through the OEA-PRJ-29 and the Regular Associateship
grants.

\bibliographystyle{spr-mp-nameyear-cnd}

\end{document}